\documentclass[twocolumn,tighten]{aastex701}
\shorttitle{EMISSIONS FROM AN EXPANDING LOOP}
\shortauthors{Dai et al.}
\usepackage{amsmath}

\begin{document}

\title{Influence of Cross-sectional Expansion on Coronal Emissions from a Radiatively Cooling Solar Flare Loop}

\author[orcid=0000-0001-9856-2770,gname=Yu,sname=Dai]{Yu~Dai}
\affiliation{School of Astronomy and Space Science, Nanjing University, Nanjing 210023, People's Republic of China}
\affiliation{Key Laboratory of Modern Astronomy and Astrophysics (Nanjing University), Ministry of Education, Nanjing 210023, People's Republic of China}
\email{ydai@nju.edu.cn}

\author[orcid=0009-0004-4898-638X,gname=Shihan,sname=Li]{Shihan~Li}
\affiliation{School of Astronomy and Space Science, Nanjing University, Nanjing 210023, People's Republic of China}
\email{lishh@smail.nju.edu.cn}

\author[gname=Wenlong,sname=Tang]{Wenlong~Tang}
\affiliation{School of Astronomy and Space Science, Nanjing University, Nanjing 210023, People's Republic of China}
\email{502024260011@smail.nju.edu.cn}

\author[gname=Zhen,sname=Li]{Zhen~Li}
\affiliation{School of Astronomy and Space Science, Nanjing University, Nanjing 210023, People's Republic of China}
\affiliation{Key Laboratory of Modern Astronomy and Astrophysics (Nanjing University), Ministry of Education, Nanjing 210023, People's Republic of China}
\email{lizhen@nju.edu.cn}
		
\author[orcid=0000-0002-4978-4972,gname=Mingde,sname=Ding]{Mingde~Ding}
\affiliation{School of Astronomy and Space Science, Nanjing University, Nanjing 210023, People's Republic of China}
\affiliation{Key Laboratory of Modern Astronomy and Astrophysics (Nanjing University), Ministry of Education, Nanjing 210023, People's Republic of China}
\email{dmd@nju.edu.cn}

\begin{abstract}
Loop-aligned hydrodynamic modelings help better understand the thermodynamic evolution of flaring plasma confined in solar flare loops. Conventional loop modelings typically assume a uniform loop cross section. With a variation of the cross section taken into account, in this work we carry out both analytical and numerical modelings of the radiative cooling in a solar flare loop. It is found that a cross-sectional expansion with height can efficiently suppress the draining of loop material from the corona while not significantly affecting the decrease of loop temperature. Reflected to the loop energetics, the coronal part of the loop cools more dominantly by radiation, and more importantly, the loop radiative outputs are shifted toward lower temperatures. These findings pose important physical implications for extreme-ultraviolet (EUV) late-phase emissions discovered in some solar flares. The late-phase loops in these flares are believed to bear a more notable cross-sectional expansion owing to their longer lengths. Compared with the main-phase loops, the late-phase loops would emit more effectively at middle temperatures, which could, to a certain degree, mitigate the severe heating requirement for the production of a prominent warm coronal  late-phase peak. In addition, the cross-sectional expansion also affects the shape of the emission lights curves, causing a sharper decay after the emission peak. Such an emission pattern has been validated with the observations of an EUV late-phase flare, and could serve as a potential diagnostic tool to judge the degree of loop cross-sectional expansion in an extended flare dataset.
\end{abstract}


\section{Introduction} 
Solar flares are one of the most energetic phenomena taking place in the solar atmosphere. The energy powering a solar flare stems from coronal magnetic fields, and magnetic reconnection plays a crucial role in energy release and conversion \citep{Parker1963,Priest2002}, which consequently give rise to electromagnetic radiation enhancements over a wide range of wavelengths, solar energetic particles (SEPs), and sometimes a coronal mass ejection (CME) propagating into the interplanetary space \citep{Reames1999,Lin2000,Fletcher2011,Benz2017}. 

According to the standard solar flare model \citep{Carmichael1964,Sturrock1966,Hirayama1974,Kopp1976}, the flare radiation originates from reconnection produced post-flare loops (flare loops for short) that are subject to an impulsive heating. When the heating of a flare loop is terminated, the loop turns into a cooling, which can be regarded as a hydrodynamic process involving circulations of both energy and mass between different atmospheric layers along the loop \citep{Reale2014}. The early stage cooling of the loop is dominated by thermal conduction; a large amount of heat flux (sometimes in combination with a beam of non-thermal electrons) travels down to the transition region (TR) and chromosphere, with the excess energy driving an evaporative flow to fill the corona \citep{Neupert1968,Antiochos1978}. As the loop density increases and temperature decreases, radiation gradually takes over. During the radiative cooling stage, the inhomogeneity of radiative losses alters the pressure structure along the loop, which in turn induces an enthalpy flow falling back from the corona to compensate for the much stronger TR radiation \citep{Bradshaw2005,Bradshaw2010}. Accompanying with the evolution of the temperature and density, the flare loops brighten up in soft X-ray (SXR) and extreme-ultraviolet (EUV) wavelengths, with the emission peaks occurring sequentially in an order of decreasing temperatures \citep{Chamberlin2012}.

Due to their pronounced geo-effectiveness on Earth's ionosphere and thermosphere \citep{Strickland1995,Bekker2025}, a proper characterization of the flare SXR and EUV emissions in terms of a loop-aligned hydrodynamic modeling is of particular importance. Conventional loop modeling studies, both analytical and numerical, typically make some simplifications. For example, the loop is assumed to have a uniform cross section, and the thermal conduction along the loop is dominated by Coulomb collisions \citep{Spitzer1953}. Under such assumptions, the modeling usually yields a loop cooling time too short to account for the observed duration of the loop emissions \citep{Ugarte-Urra2006,Warren2010}. Such discrepancy can be largely reconciled by a suppression of the thermal conduction, which can be realized either by incorporating new conduction conditions such as turbulent scattering \citep{Bian2016,Bian2018} and free streaming \citep{Klimchuk2008,Zhu2018}, or simply by altering the loop geometry. In particular, if the loop cross section expands with height, the downward cross-sectional contraction would effectively block the transport of heat flux, hence leading to a prolonged loop cooling \citep{Antiochos1976,Reep2022}.

In addition to emission duration, the intensity of the loop emissions is another important issue that deserves our further attention. Using EUV observations with the \emph{Solar Dynamics Observatory} \citep[\emph{SDO};][]{Pesnell2012}, \citet{Woods2011} discovered an ``EUV late phase"  in some solar flares, which is manifested as a second peak in the warm coronal emissions (e.g., the \ion{Fe}{16} 335 {\AA} emission at $\sim$3 MK) several tens of minutes to a few hours after the main flare peak. Imaging observations confirm that the late-phase emission originates from a set of longer and higher flare loops other than the main flare loops \citep{Liu2013,Dai2013,Sun2013,Masson2017}, and statistical studies reveal that a large fraction of late-phase flares exhibit an extremely large late-phase peak that is even stronger than the main-phase peak \citep{Chen2020,Ornig2025}. Although a long-lasting cooling in the long late-phase loops can naturally explain the delayed occurrence of the late-phase peak \citep{Liu2013,Masson2017,Dai2018b,Chen2023}, its high intensity seems to impose an extreme energy requirement for the initial flare heating of the late-phase loops \citep{Dai2018}. It has also been suggested that an additional persistent heating may promote the late-phase peak to a high level \citep{Liu2015,Zhou2019}, but such an additional heating is not common among late-phase loops.

By analyzing the heating and cooling histories of an EUV late-phase flare, \citet{Li2024} have recently proposed that in addition to an intense early heating, a cross-sectional expansion of the late-phase loops can also make a contribution to the observed extremely large late-phase peak by sustaining the loop density for a long time. Motivated by this argument, in this work we carry out both analytical and numerical hydrodynamic modelings of the radiative cooling in an expanding flare loop. We focus on the radiative cooling since the majority of flare emissions are released during this stage. As we will show, the inclusion of a cross-sectional expansion can essentially alter the partition of the radiative outputs between different temperatures, consequently influencing the intensity and shape of the emission light curves.  The rest of the paper is organized as follows. In Section \ref{sec02} we construct analytical solutions of the radiative cooling, based on which the effect of loop cross-sectional expansion is explored. The analytically predicted physical picture is numerically verified in Section \ref{sec03}, and observationally validated in Section \ref{sec04}. Finally, the results are discussed and conclusions are drawn in Section \ref{sec05}.

\section{Analytical Modeling\label{sec02}}

\subsection{Model and Solutions for Radiative Cooling}
Since we model a solar flare loop as a coherent coronal structure following a ``rigid" magnetic flux tube, the dynamic evolution of plasma inside the loop is governed by the one-dimensional, time-dependent hydrodynamic equations that involve the conservation of mass, momentum, and energy, i.e., 
\begin{equation}
\frac{\partial\rho}{\partial t}+\frac{1}{A}\frac{\partial}{\partial s}\left(A\rho v\right)=0,\label{eoc1}  
\end{equation}
\begin{equation}
\frac{\partial}{\partial t}\left(\rho v\right)+\frac{1}{A}\frac{\partial}{\partial s}\left(A\rho v^2\right)=-\frac{\partial p}{\partial s}+\rho g_{\scriptscriptstyle\parallel},\label{eom1}
\end{equation}
\begin{equation}
\begin{aligned}
 \frac{\partial \varepsilon}{\partial t}+\frac{1}{A}\frac{\partial}{\partial s}\left[A(\varepsilon+p)v\right]&=-\frac{1}{A}\frac{\partial }{\partial s}\left(A F_{C}\right)\\
&+Q_{H}+Q_{R}+\rho g_{\scriptscriptstyle\parallel}v,\label{eoe1} 
\end{aligned}
\end{equation}
where $\rho=n m_i$ is the mass density with $n$ being the electron number density and $m_i$ the average ion mass, $A$ is the loop cross-sectional area, $v$ is the bulk  flow velocity, $p$ is the gas pressure,  $g_{\scriptscriptstyle\parallel}$ is the gravitational acceleration parallel to the loop, $\varepsilon=p/(\gamma-1)+\rho v^2/2$ is the total energy density with $\gamma$ being the specific heat ratio, $F_C=-\kappa\partial T/\partial s$ is the loop-aligned heat flux with $T$ being the temperature and $\kappa$ the thermal conduction coefficient, $Q_H$ is the volumetric heating rate, and $Q_R=-n^2\Lambda(T)$ is the optically thin radiative loss rate with $\Lambda(T)$ representing the radiative loss function. Using Equations (\ref{eoc1}) and (\ref{eom1}) to eliminate the kinetic energy related terms, Equation~(\ref{eoe1}) reduces to
\begin{equation}
\begin{aligned}
\frac{\partial}{\partial t}\left(\frac{p}{\gamma-1}\right)+\frac{1}{A}\frac{\partial }{\partial s}\left(AF_{E}\right)&=-\frac{1}{A}\frac{\partial }{\partial s}\left(AF_{C}\right)\\
&+v\frac{\partial p}{\partial s}+Q_{H}+Q_{R},\label{eoe1b}
\end{aligned}
\end{equation}
where $F_E=\gamma pv/(\gamma-1)$ denotes the enthalpy flux. \added{Further applying the equation of state for fully ionized plasma $p=2k_BnT$  (with $k_B$ being the Boltzmann constant), the above equations can be solved with respect to time $t$ and loop-aligned coordinate $s$. }

For a flare loop experiencing its radiative cooling stage, the flare heating has been turned off ($Q_H=0$) and thermal conduction becomes negligible ($F_C\approx0$).  With a still high enough loop temperature, the pressure scale height of the loop is so large (compared with its geometrical height) that the effect of gravity is negligible ($\rho g_{\scriptscriptstyle\parallel}\approx0$) and any flows inside the loop should be driven by a pressure gradient along the loop. Meanwhile, the direction of the gradient-driven flows is such that they tend to smooth out the pressure non-uniformity. In this sense, the loop is largely isobaric ($\partial p/\partial s\approx0$) and the flows are assumed to be subsonic ($\rho v^2\ll p$). Under such assumptions, the above governing equations can be substantially simplified to
\begin{equation}
\displaystyle \frac{\partial n}{\partial t}+\frac{1}{A}\frac{\partial}{\partial s}\left(Anv\right)=0,\label{eoc2} 
\end{equation}
\begin{equation}
\frac{\partial p}{\partial s} = 0,\label{eom2}
\end{equation}
\begin{equation}
\frac{\partial p}{\partial t}+(\gamma-1)\left[\frac{1}{A}\frac{\partial}{\partial s}\left(\frac{A\gamma pv}{\gamma-1}\right)+n^2\Lambda(T)\right]=0.\label{eoe2}
\end{equation}
As revealed in Equation (\ref{eoe2}), the evolution of internal energy in a radiatively cooling loop is dominantly determined by radiation and enthalpy conduction. 

To construct analytical solutions for the radiative cooling,  the optically thin radiative loss function is conventionally approximated by a single power-law form of
\begin{equation}
\Lambda(T)=\chi_0 T^{-l},\label{rlf}
\end{equation}
where $l\ (l>0)$ characterizes the slope of the function and $\chi_0$ is the coefficient of proportionality. \added{Inserting Equation (\ref{rlf}) to Equation (\ref{eoe2}), it gives an analytical estimate of the radiative cooling timescale as
\begin{equation}
\tau_{R0}=\frac{2k_BT_0^{l+1}}{(\gamma-1)\chi_0 n_0},\label{taur0}
\end{equation}
where $T_0$ and $n_0$ are the looptop temperature and density at the start of the radiative cooling.} In addition, we assume a semi-circular geometry of the loop. For symmetry, we only need to consider half of the loop, whose length is $L$ measured from either footpoint of the loop to its apex.

\added{\citet{Antiochos1980} and \citet{Cargill1995} proposed a separation of variables to solve the above equations. By extending their original formalism, we have recently constructed full form analytical solutions of the radiatively cooling loop, casting them as
\begin{equation}
T(s,t)=\xi^{1/(l+2)}(1+\eta t)^{\delta/[(l+1)\delta-1]}T_{0}\label{solt}
\end{equation}
and
\begin{equation}
n(s,t)=\xi^{-1/(l+2)}(1+\eta t)^{1/[(l+1)\delta-1]}n_{0},\label{soln}
\end{equation}
where $\xi$ is a function of $s$ that characterizes the temperature structure along the loop, $\eta$ quantifies the cooling rate of the loop, and $\delta$ is an index relating the evolutions of temperature and density by a scaling relation of $T\sim n^{\delta}$. 

Using the scaling index $\delta$, $\eta$ can be explicitly expressed as 
\begin{equation}
\eta=-\frac{(l+1)\delta-1}{(\delta-\gamma+1)\tau_{R0}},\label{etadelta}
\end{equation}
and $\xi$ can be implicitly formulated  by
 \begin{equation}
\beta_r\left(\xi;\frac{l+3}{l+2},\frac{\gamma}{(l+2)(\delta-\gamma+1)}\right)=\frac{\int_0^sA(s)ds}{\bar{A}L},\label{tp1}
\end{equation}
where $\beta_r$ denotes the regularized incomplete Beta function, and $\bar{A}=\int_0^LA(s)ds/L$ is the average cross-sectional area of the loop. Further applying the equation of continuity, the bulk velocity is then solved as
\begin{equation}
\begin{aligned}
v(s,t)&=-\left[B\left(\frac{l+3}{l+2},\frac{\gamma}{(l+2)(\delta-\gamma+1)}\right)\right]^{-1}\\
&\ \times(l+2)\xi^{1/(l+2)}\left(1-\xi\right)^{{\gamma}/{(l+2)(\delta-\gamma+1)}}\\
&\ \times(1+\eta t)^{-1}\frac{(\gamma-1)\chi_0 n_0\bar{A}L}{2\gamma k_BT_0^{l+1}A(s)},\label{solv}
\end{aligned}
\end{equation}
with $B$ representing the complete Beta function. For more details on the solution construction, please refer to \citet[][their Appendix B.2]{Li2024}.}

\subsection{Effect of Loop Cross-sectional Expansion}
\subsubsection{Physical Meaning of the Scaling Index $\delta$}
Since the $T$--$n$ scaling index $\delta$ has not be determined so far,  Equations (\ref{solt}) and (\ref{soln}) may represent an infinite set of possible solutions corresponding to different values of $\delta$. Nevertheless, it does not always guarantee the existence of valid loop solutions for an arbitrarily specified $\delta$. \added{Following the discussion presented in \citet{Li2024}, a temperature profile of both mathematical and physical significances requires that}
\begin{equation}
\delta>\frac{l+3}{l+2}\gamma-1,\label{cvdelta}
\end{equation}
which gives a critical value for $\delta$, below which no valid solutions for the radiative cooling could exist. Note that the lower bound of $\delta$, $\gamma-1$, occurring only when $l\to\infty$, corresponds to an adiabatic draining of material from the corona to compensate for the extremely large radiative losses in the TR. Further relating Equation (\ref{cvdelta}) to Equation (\ref{etadelta}), it is easily verified that $\eta<0$ for any positive $l$, as long as $\gamma$ is not less than 4/3 (a condition naturally satisfied in the solar atmosphere where the specific heat ratio $\gamma$ is assumed to be 5/3).

\begin{figure*}
\epsscale{0.85}
\plotone{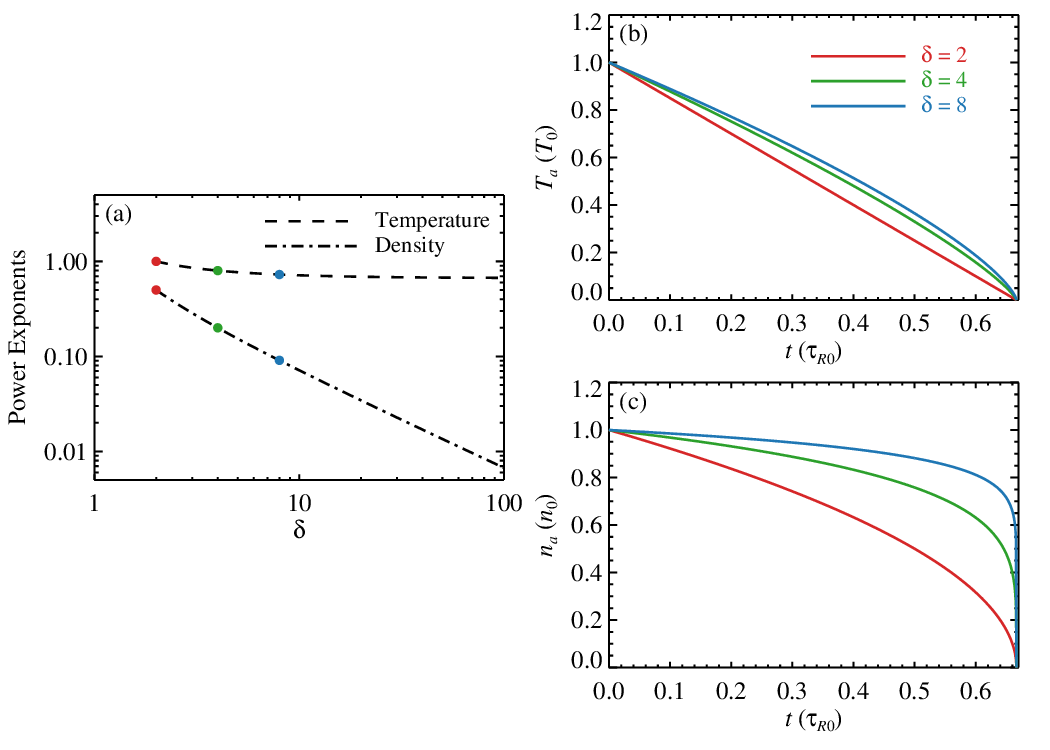}
\caption{Analytical solutions for radiative cooling. Panel (a) plots the variations of the power exponents on $(1+\eta t)$ versus $\delta$ for both temperature (dashed curve) and density (dashed--dotted curve), where the color-filled circles highlight the results evaluated at $\delta=2$ (red), $\delta=4$ (green), and $\delta=8$ (blue), respectively. Panels (b) and (c) display the temporal profiles of the looptop temperature and density for three different values of  $\delta$, which are color-coded according to panel (a). In computing the analytical solutions, the values of $\l$ and $\gamma$ are set to be 1/2 and 5/3, respectively.\label{fig01}}
\end{figure*}

Within the permitted data range constrained by Equation (\ref{cvdelta}), a higher value of $\delta$ in the $T$--$n$ scaling relation implies either a more pronounced temperature drop with respect to density, or a less prominent density decrease with respect to temperature. \added{To clarify which situation is more appropriately revealed by our analytical solutions, in Figure \ref{fig01}(a) we plot the variations of the power exponents on $(1+\eta t)$ versus $\delta$ for both temperature (Equation (\ref{solt})) and density (Equation (\ref{soln})).} Here, we set $l=1/2$ and $\gamma=5/3$ as commonly adopted in other studies (as well as the following analytical calculations in this work), by which $\eta=-3/2\tau_{R0}$ (corresponding to an overall radiative cooling time of $2\tau_{R0}/3$) in all cases irrespective of the specification of $\delta$. \added{It is seen that both power exponents decrease with the increase of $\delta$. By comparison, the exponent for density evolution varies much more significantly with $\delta$, whereas that for temperature evolution just depends weakly on $\delta$. Reflected to the temporal evolutions of the looptop temperature $T_a$ and density $n_a$, which are shown in Figures \ref{fig01}(b) and (c) for three different values of $\delta$, it is found that with the increase of $\delta$, the shape of the temporal profiles becomes increasingly convex, and more importantly, the pattern of convexness is much more prominent for the temporal evolution of density.} Therefore, under a fixed radiative loss function, the increase of $\delta$ predominately means a suppression of mass draining from the corona, which would last for a long time during the radiative cooling stage. Physically it could be realized by an expansion of the loop cross section with height, or in other words, a downward cross-sectional contraction, which acts like a bottle neck to block the draining of coronal material back toward the TR\@. 

Using typical radiative loss functions for the solar circumstance, previous numerical studies have shown that $\delta\approx2$ for a uniform loop cross section \citep{Serio1991,Jakimiec1992,Bradshaw2010}. Here, we adopt it as the baseline value for $\delta$, with any higher value of $\delta$ reflecting a certain degree of cross-sectional expansion.

\subsubsection{Radiation and Enthalpy Conduction}

To explore the effect of cross-sectional expansion on loop thermodynamics, we further calculate the energy terms of radiation and enthalpy conduction based on the analytical solutions. Considering a variation of the loop cross section, the total radiative loss over the half loop $\mathcal{R}_L$ can be derived from an integral of
\begin{equation}
\begin{aligned}
\mathcal{R}_L(t)&=-\int_0^LA(s)n^2\Lambda(T)ds=-\frac{\delta+1}{\delta-\gamma+1}\\
&\  \times(1+\eta t)^{-(l\delta-2)/[(l+1)\delta-1]}\chi_0 n_0^2T_0^{-l}\bar{A}L.\label{solrl}
\end{aligned}
\end{equation}
Meanwhile, the cross-sectionally integrated enthalpy flow $\mathcal{F}_E=AF_E$ at any loop coordinate is given by
\begin{equation}
\begin{aligned}
\mathcal{F}_E(s,t)&=-\left[B\left(\frac{l+3}{l+2},\frac{\gamma}{(l+2)(\delta-\gamma+1)}\right)\right]^{-1} \\
&\ \times(l+2)\xi^{1/(l+2)}\left(1-\xi\right)^{{\gamma}/{(l+2)(\delta-\gamma+1)}}\\
&\ \times(1+\eta t)^{-(l\delta-2)/[(l+1)\delta-1]}\chi_0 n_0^2T_0^{-l}\bar{A}L.\label{solfe}
\end{aligned}
\end{equation}

The minus sign on the right hand side of Equation (\ref{solfe}) indicates a downward transporting enthalpy flow during the radiative cooling stage. Moving down from the loop apex to base, the magnitude of the enthalpy flow first increases from zero and then decreases back to zero, implying a redistribution of internal energy along the loop by enthalpy conduction without any net loss. The point of $-\mathcal{F}_E$ maximum, located at $\xi=(\delta-\gamma+1)/(\delta+1)$, can be mathematically defined as the interface between the coronal and TR parts of the loop, where the interface enthalpy flow $\mathcal{F}_{Ei}$ is
\begin{equation}
\begin{aligned}
\mathcal{F}_{Ei}(t)&=-\left[B\left(\frac{l+3}{l+2},\frac{\gamma}{(l+2)(\delta-\gamma+1)}\right)\right]^{-1}(l+2)\\
&\ \times\left(\frac{\delta-\gamma+1}{\delta+1}\right)^{1/(l+2)}\left(\frac{\gamma}{\delta+1}\right)^{{\gamma}/{(l+2)(\delta-\gamma+1)}}\\
&\ \times(1+\eta t)^{-(l\delta-2)/[(l+1)\delta-1]}\chi_0 n_0^2T_0^{-l}\bar{A}L.\label{solfei}
\end{aligned}
\end{equation}
By this definition, enthalpy conduction acts as an energy loss (cooling) term in the corona, with $\mathcal{F}_{Ei}$ representing the total conductive enthalpy drained from the corona, whereas in the TR, it turns into an energy gain (heating) term to power the TR radiation. Assuming a quasi-steady TR, a rough balance of energy over the TR can be established as
\begin{equation}
\mathcal{R}_t\approx\mathcal{F}_{Ei},\label{solrt}
\end{equation}
where $\mathcal{R}_t$ is the total radiation over the TR following the definition of $\mathcal{R}_L$. Using this balance relation, the total radiation over the corona $\mathcal{R}_c$ is then written as
\begin{equation}
\mathcal{R}_c=\mathcal{R}_L-\mathcal{R}_t\approx\mathcal{R}_L-\mathcal{F}_{Ei},\label{solrc}
\end{equation}
which can be readily obtained by relating Equations (\ref{solrl}) and (\ref{solfei}).

Although the areal function $A(s)$ appears in the original definition of $\mathcal{R}_L$ and $\mathcal{F}_E$, it can be eliminated in the intermediate calculations. Without the need of an explicit form of $A(s)$, here the effect of non-uniform cross section is implicitly reflected by the specification of $\delta$.

\begin{figure}
\epsscale{1}
\plotone{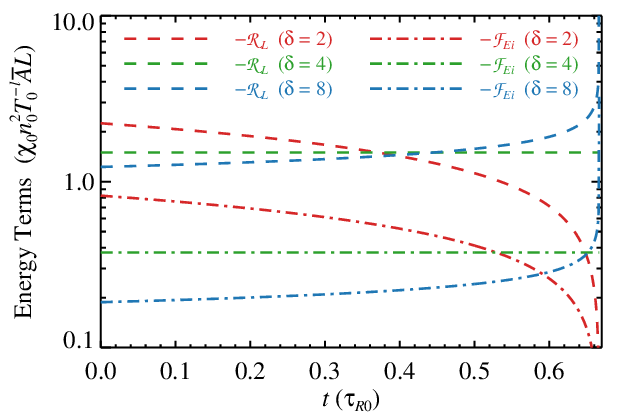}
\caption{Temporal profiles of the total loop radiation (dashed curves) and interface enthalpy flow (dashed--dotted curves)  derived from the analytical solutions (Equations (\ref{solrl}) and (\ref{solfei})) for different values of $\delta$ (discriminated with different colors).\label{fig02}}
\end{figure}

Figure \ref{fig02} shows the temporal evolutions of $-\mathcal{R}_{L}$ and $-\mathcal{F}_{Ei}$  for different values of $\delta$. Note that we add a minus sign before both quantities, so that they are plotted in absolute magnitudes. For each given value of $\delta$, the two quantities follow the same temporal evolution except for their magnitudes. This can be expected from their analytical expressions (Equations (\ref{solrl}) and (\ref{solfei})), where the power exponents on $(1+\eta t)$ are the same for both quantities. With the increase of $\delta$, it is found that the temporal trend of the evolution exhibits an obvious change: for $\delta=2$ (i.e., the case of a uniform cross section), the magnitudes of both quantities decrease with time, for $\delta=4$ (e.g., the case of a moderately expanding cross section), they keep unchanged throughout the whole radiative cooling stage, and for $\delta=8$ (e.g., the case of a strongly expanding cross section), the magnitudes increase with time instead. This change is mathematically due to a transition of the sign of the power exponent $-(l\delta-2)/[(l+1)\delta-1]$  from positive (when $\delta<4$) to negative (when $\delta>4$), which alters the monotonicity of function as the base factor $(1+\eta t)$ decreases with $t$. Physically it has an important implication for the thermodynamic evolution of a radiatively cooling flare loop. With an expansion of the loop cross section, the radiation from the loop will increasingly shift toward lower temperatures.

\begin{figure}
\epsscale{1}
\plotone{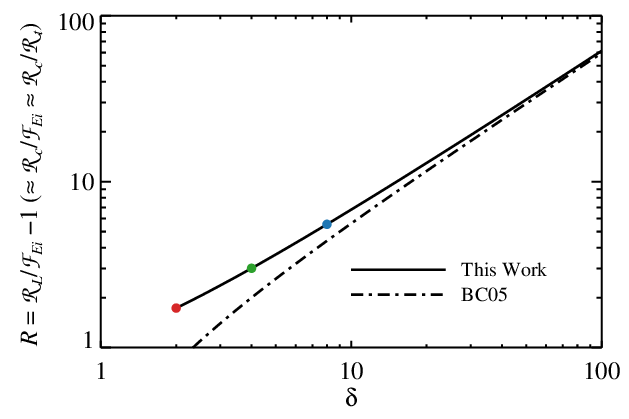}
\caption{Variation of the dimensionless ratio $R$ (=$\mathcal{R}_L/\mathcal{F}_{Ei}-1$) vs. the scaling index $\delta$. The solid curve is derived from the analytical solutions in this work, where the color-filled  circles highlight the results evaluated at $\delta=2$ (red), $\delta=4$ (green), and $\delta=8$ (blue), respectively, and the dashed--dotted curve is based on the approach in \citet{Bradshaw2005}, which is plotted here for comparison.\label{fig03}}
\end{figure}

To further ascertain the relative contributions of radiation and enthalpy conduction during the radiative cooling stage, we introduce a dimensionless ratio $R$ defined as
\begin{equation}
R=\frac{\mathcal{R}_L}{\mathcal{F}_{Ei}}-1,
\end{equation}
which is a time-invariant function of $\delta$ since both terms involving $t$ in $\mathcal{R}_L$ and $\mathcal{F}_{Ei}$ are cancelled out through a reduction of fraction. Considering the coronal part alone, $R$ denotes the ratio of radiative loss to enthalpy loss from the corona ($R\approx\mathcal{R}_c/\mathcal{F}_{Ei}$); relating the corona to TR, it further represents the ratio of radiative losses between the corona and TR ($R\approx\mathcal{R}_c/\mathcal{R}_t$).

Figure \ref{fig03} displays the variation of $R$ versus $\delta$ according to our analytical solutions (plotted in solid curve), with three additionally plotted color-filled circles highlighting the results evaluated at  $\delta=2$, $\delta=4$, and $\delta=8$, respectively (corresponding to the three cases demonstrated in Figure \ref{fig02}). As shown in the figure, the ratio $R$ starts from 1.73 (for $\delta=2$), and increases monotonically with $\delta$, suggesting an increasingly dominant role of radiative loss over enthalpy conduction in removing energy from the corona. Such pattern can be readily understood in terms of an increasing degree of loop cross-sectional expansion. By suppressing the mass draining, the cross-sectional expansion also reduces the coronal enthalpy loss, and so the loop internal energy must be more removed by radiation. 

By artificially presupposing a corona-TR interface and averaging the energy equation over the corona and TR separately, \citet[][hereafter BC05]{Bradshaw2005} also quantitatively established a relation between $R$ and $\delta$, which is rewritten here as
\begin{equation}
R=\frac{\delta-\gamma+1}{\gamma}\label{bc05}
\end{equation}
after an adaptation (and typo correction) of its original form (Equation (21) in their paper). In Figure \ref{fig03} we over-plot the results based on their approach (dashed--dotted curve) for comparison. It is found that our analytically derived curve exhibits a good agreement with the BC05 curve at large values of $\delta$. At $\delta=100$, for instance, the discrepancy between the two curves has fallen below a level of 3\%.  For our approach, using the analytical approximation of the Beta function  in Equation (\ref{solfei}), it is readily verified that $R\approx\delta/\gamma$ as $\delta\rightarrow\infty$, exactly equivalent to the corresponding asymptotical expression of Equation (\ref{bc05}). Nevertheless, we note that the BC05 approach is based on the assumption of an ideally isothermal corona, which holds only for a very large $\delta$ according to our analytical calculation, but will break when $\delta$ is relatively small. Therefore, it is not surprising that our curve gradually deviates from the BC05 curve at smaller values of $\delta$. In this regime, our analytical approach should present a more robust quantification of the loop thermodynamics.

\subsubsection{Synthetic Emission Light Curves}
The loop thermodynamic evolution can be observationally diagnosed via the emission light curves in EUV wavelengths. To this end, we use the loop quantities derived from our analytical solutions to forward synthesize loop emissions in some optically thin EUV lines and passbands. For an emission line, the irradiance from a localized part of loop, as observed from a near-Earth perspective, is given by
\begin{equation}
I_{\mathrm{line}}=G(T)n^2z_d\frac{S}{4\pi d_s^2}\ \  (\mathrm{erg\ cm^{-2}\ s^{-1}}),\label{synline}
\end{equation}
where $G(T)$ is the contribution function of the line, which can be calculated with the CHIANTI atomic database \citep{DelZanna2021,Dufresne2024}, $S$ and $z_d$ are the projection area and line-of-sight (LOS) depth of the emitting volume, respectively, and $d_s$ is the Sun-Earth distance. Meanwhile, the filtergram intensity from the same emitting region, as recorded by a passband of the Atmospheric Imaging Assembly \citep[AIA;][]{Lemen2012} on board \emph{SDO}, is modeled as
\begin{equation}
I_{\mathrm{AIA}}=K(T)n^2z_d\frac{S}{S_{\mathrm{pix}}}\ \ (\mathrm{DN\ s^{-1}}),\label{synaia}
\end{equation}
where $K(T)$ is the temperature response function of the passband, which can be obtained by calling the Solar Software \citep[SSW;][]{Freeland1998} routine \texttt{aia\_get\_response.pro}, and $S_{\mathrm{pix}}$ is a normalization factor that represents the projection area on the Sun covered by one AIA pixel. 

We consider a flare loop during its radiative cooling stage. The loop has a half-length of 50 Mm, and the initial looptop temperature is set to be 8 MK, a typical temperature at the start of radiative cooling \citep{Dai2018}. Since the transition point from conductive cooling to radiative cooling is in proximity to an equilibrium state \citep{Reale2014}, we put the above loop parameters into a numerical hydrostatic modeling, which yields an initial looptop density of $\sim1.4\times10^{10}$ cm$^{-3}$ (see Section \ref{sec02}), largely consistent with the equilibrium loop scaling law proposed by \citet{Rosner1978}. Further adopting $l=1/2$ and $\chi_0=1.55\times10^{-19}$~erg~cm$^3$~s$^{-1}$~K$^{1/2}$ for the radiation function (see Equation~(\ref{rlf})), we obtain an overall radiative cooling time ($=2\tau_{R0}/3$) of 2880 s for the loop (see Equation (\ref{taur0})). 

As the loop cools down from a moderately hot coronal temperature to lower temperatures, we synthesize loop emissions in two iron EUV lines of \ion{Fe}{16} 335~{\AA} and \ion{Fe}{9} 171~{\AA}, and two AIA EUV passbands of AIA 335~{\AA} and AIA 171~{\AA}. For the two EUV lines, their contribution functions peak around  2.7~MK and 0.8~MK, reflecting emissions from warm and cool coronae, respectively. In spite of a broader temperature response, in each considered AIA passband, the dominant contributing ion is the same as the emitting ion of the corresponding iron line.

The synthetic loop emissions are sampled in a local volume at the loop apex, which has a projection area of $0.43\times0.43$~Mm$^2$ and LOS depth of $1.5$ Mm. Here $S$ is set to equal the size of one AIA pixel so that it can be cancelled out in Equation (\ref{synaia}), and $z_d$ is consistent with the width of a typical coronal loop \citep{Mandal2024,Vasantharaju2025}.

\begin{figure}
\epsscale{1}
\plotone{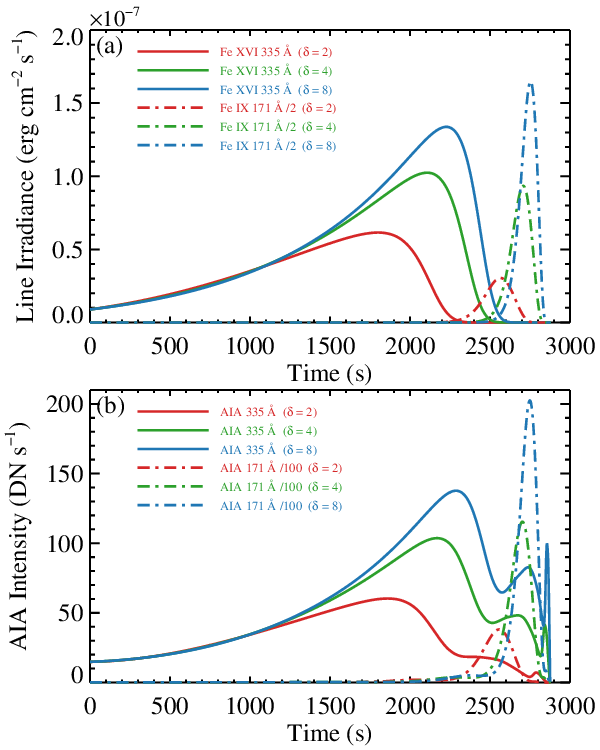}
\caption{Light curves of the apex region of a radiatively cooling flare loop, which are synthesized with the analytical solutions for  different values of $\delta$. The upper and lower panels display the synthetic light curves in two iron lines and two AIA passbands, respectively. The meaning of the color and style for each curve is detailed in the legend.\label{fig04}}
\end{figure}

Figure \ref{fig04} plots the synthetic emission light curves of the loop apex region for $\delta=2$, $\delta=4$, and $\delta=8$, respectively, which reveal a significant impact of cross-sectional expansion on the loop emissions. First, the increase of $\delta$  leads to a remarkable elevation of the peak intensity for both the iron lines and the AIA passbands. Second, the occurrence of the peak is increasingly delayed. And third, the intensity after the peak exhibits a sharper decay. The latter two features are more prominent for \ion{Fe}{16}/AIA 335 {\AA}. In passing, we note that the AIA 335 {\AA} light curves also exhibit secondary bumps in addition to the main emission peak, which are due to the broad temperature coverage of this passband.

\begin{deluxetable}{lccccchccccc}
\tablecaption{Quantities Relevant to the Peak of Synthetic Looptop Emissions for the Analytical Solutions\label{table01}}
\tablehead{
\colhead{} & \multicolumn{5}{c}{Peak of \ion{Fe}{16}/AIA 335 {\AA}} & \colhead{} & \multicolumn{5}{c}{Peak of \ion{Fe}{9}/AIA 171 {\AA}}\\
\cline{2-6} \cline{8-12}
\colhead{} & \colhead{$I_{\mathrm{peak}}$\tablenotemark{\scriptsize a}} & \colhead{$t_{\mathrm{peak}}$} & \colhead{$t_{\mathrm{decay}}$} & \colhead{$T_a$} & \colhead{$n_a$} & \colhead{} & \colhead{$I_{\mathrm{peak}}$\tablenotemark{\scriptsize a}} & \colhead{$t_{\mathrm{peak}}$}  & \colhead{$t_{\mathrm{decay}}$} & \colhead{$T_a$} & \colhead{$n_a$}\\
\colhead{} & \colhead{} & \colhead{\footnotesize (s)} &\colhead{\footnotesize (s)} & \colhead{\footnotesize(MK)} & \colhead{\footnotesize (cm$^{-3}$)} & \colhead{} & \colhead{} & \colhead{\footnotesize (s)} & \colhead{\footnotesize (s)} & \colhead{\footnotesize (MK)} & \colhead{\footnotesize(cm$^{-3}$)}
}
\startdata
& \multicolumn{11}{c}{Fe line irradiance}\\
$\delta=2$  & $6.15\times10^{-8}$  & 1800 & 310 & 3.00 & $8.57\times10^{9}$ &   & $6.09\times10^{-8}$ & 2566 & 95 & 0.867 & $4.61\times10^{9}$\\
$\delta=4$  & $1.02\times10^{-7}$  & 2107 & 237 & 2.79 & $1.08\times10^{10}$ &   & $1.87\times10^{-7}$ & 2708 & 63 & 0.832 & $7.95\times10^{9}$\\
$\delta=8$  & $1.34\times10^{-7}$  & 2227 & 207 & 2.71 & $1.22\times10^{10}$ &   & $3.30\times10^{-7}$ & 2754 & 50 & 0.812 & $1.05\times10^{10}$\\
\hline
& \multicolumn{11}{c}{AIA passband intensity}\\
$\delta=2$  & 60.3 & 1860 & 333 & 2.83 & $8.33\times10^{9}$ & & $3.81\times10^{3}$ & 2558 & 97 & 0.890 & $4.67\times10^{9}$\\
$\delta=4$  & 104  & 2169 & 267 & 2.61 & $1.06\times10^{10}$ & & $1.15\times10^{4}$ & 2702 & 66 & 0.855 & $8.01\times10^{9}$\\
$\delta=8$  & 138 & 2287 & 248 & 2.53 & $1.21\times10^{10}$ &  & $2.02\times10^{4}$ & 2749 & 52 & 0.836 & $1.06\times10^{10}$
\enddata
\tablenotetext{\scriptsize a}{\footnotesize Values are listed in units of erg cm$^{-2}$ s$^{-1}$ for the Fe line irradiance, or DN s$^{-1}$ for the AIA passband intensity.}
\end{deluxetable} 

The above emission patterns are quantitatively illustrated in Table \ref{table01}, which lists the quantities relevant to the peak of the synthetic light curves, including the peak intensity ($I_{\mathrm{peak}}$), peak time ($t_{\mathrm{peak}}$), decay time ($t_{\mathrm{decay}}$, defined as the time it takes for the emission intensity to decrease from its peak to the half maximum level), and looptop temperature and density at the peak time. For each iron line or AIA passband, it is interesting to note that the temperature at the time of emission peak exhibits a tendency of decrease with the increase of $\delta$, asymptotically approaching the peak formation temperature of the corresponding contribution or response function. 

According to  Equations (\ref{synline}) and (\ref{synaia}), the rise of loop emission  during the radiative cooling stage is the result of a competition between the increase of the contribution/response function and the decrease of the density. If the loop material is drained at a relatively high rate (e.g., for the case of a uniform loop cross section), the effect of the density decrease may win out soon, so that the emission intensity would peak at an earlier time and turn into a decay when the contribution/response function is still increasing. On the other hand, if the mass draining is strongly suppressed (e.g., for the case of a strong cross-sectional expansion), the emission intensity will keep increasing until the maximum of the contribution/response function is reached. In such situation, both the density factor (a higher level of the density) and the temperature factor (a larger value of the contribution/response function) work together, leading to a higher peak intensity.

As to the decay stage, for the uniform cross section, the still increasing contribution/response function will mitigate the drop of the emission intensity at first, hence leading to a somewhat gradual decay. By comparison, for the strong cross-sectional expansion, the contribution/response function will start to decrease immediately after the emission peak, so the emission intensity will experience a sharper decay.

\section{Numerical Verification\label{sec03}}
To verify the physical picture revealed by the simplified analytical modeling, we further carry out numerical simulations with the publically available HYDrodynamic and RADiation code \citep[HYDRAD;][]{Bradshaw2003,Bradshaw2013}. HYDRAD is a light-weight and high-efficiency numerical code that solves the field-aligned hydrodynamic equations in coronal loops. HYDRAD considers a two-fluid plasma, but in practice the plasma can be forced to evolve as a single fluid  by artificially shortening the inter-species collisional time scale. In this way, the governing equations are largely the same as those presented in Equations~(\ref{eoc1}--\ref{eoe1}).

We numerically investigate the cooling of a flare loop with initial conditions as close as possible to the above analytically modeled loop. The loop has a full-length of 100 Mm (HYDRAD deals with full loops rather than half loops in analytical considerations). A chromosphere of a thickness of 2 Mm and a uniform temperature of $2\times10^4$ K is added to each leg of the loop, slightly extending the total computation domain to $2L=104$ Mm. The loop is initially maintained by a temporally steady and spatially uniform background heating. By setting $\partial/\partial t=0$ and $v=0$, we first integrate the hydrostatic equations to construct the initial equilibrium temperature and density profiles along the loop. Here, we consider gravitational stratification, and adopt a piecewise power-law formed function \citep{Klimchuk2008}, \added{by which we can more realistically mimic the circumstances in the solar corona.} By iteratively adjusting the heating rate and footpoint density, the loop finally attains a temperature of 8 MK at the apex, where the density is $1.38\times10^{10}$ cm$^{-3}$ for the case of a uniform cross section.

Starting with the initial temperature and density profiles, we turn off the background heating, and then trace the hydrodynamic evolution of the loop as it cools down. To avoid the occurrence of negative temperatures within the loop, the optically thin radiation function is set to be zero once the temperature drops below $2\times10^4$ K. As to the variation of the loop cross section, we adapt an explicitly distance-dependent areal function from \citet{Cargill2022} and rewrite it as
\begin{equation}
A(s)=A_b\left[1+(\Gamma-1)\sin^2\left(\frac{\pi s}{2L}\right)\right],
\end{equation}
where the factor $\Gamma\ (\Gamma\ge1)$ denotes the ratio of cross-sectional areas between the loop apex and base. The larger the value of $\Gamma$, the stronger the expansion of the cross section. When $\Gamma\to\infty$, this areal function in fact describes a closed flux tube generated by a line dipole \citep{Antiochos1980}. In the numerical modeling, a caveat should be pointed out that the initial temperature and density profiles for all cases are chosen the same as those derived for the uniform cross section ($\Gamma=1$). Although being convenient for a quantitative comparison between different numerical cases, this treatment is not physically self-consistent, because a non-uniform cross section may considerably change the temperature and density structures of a coronal loop \citep{Cargill2022,Dai2024}. Nevertheless, at the start of the radiative cooling, the loop temperature is high enough that the sound travel time scale is significantly shorter than the instantaneous radiative cooling time scale \citep{Cargill2013}. As a result, a mass flow is quickly built up in the loop, which in turn adjusts the loop quantities to adapt to the loop geometry accordingly. In the sense, the specification of the initial conditions does not significantly affect the following hydrodynamic evolution of the loop. 

\begin{figure*}
\epsscale{0.8}
\plotone{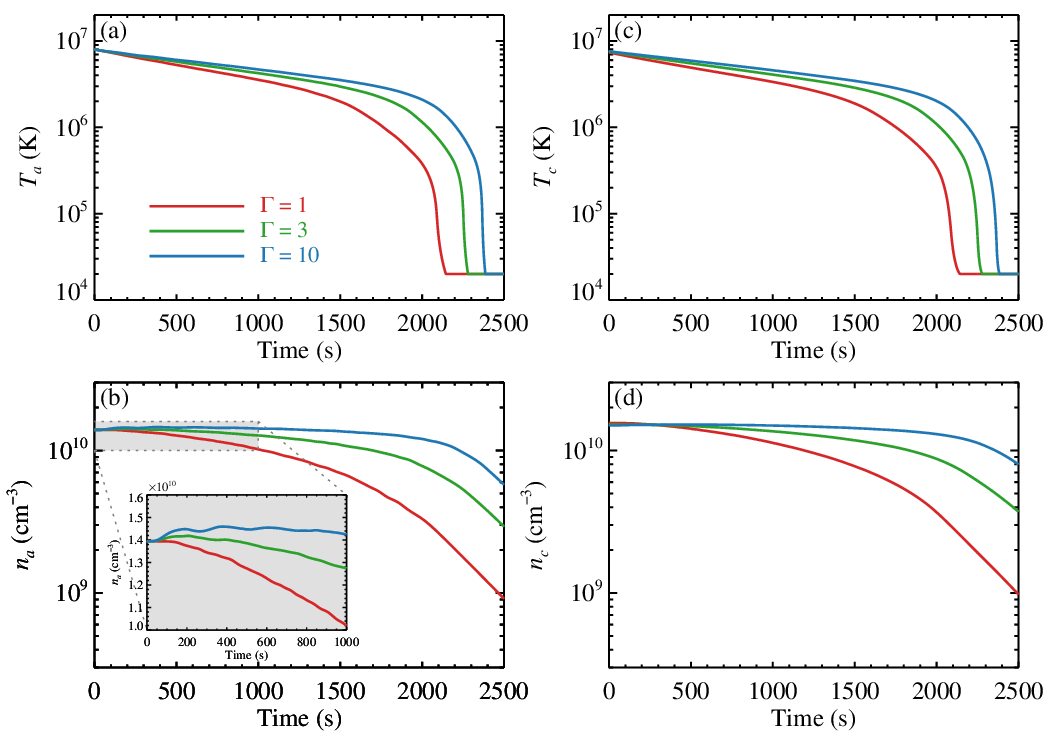}
\caption{Temporal profiles of loop quantities derived from the HYDRAD numerical simulations for different values of $\Gamma$ (discriminated with different colors). The left panels show the evolutions the looptop temperature (a) and density (b), and the right panels display the evolutions of the average coronal temperature (c) and density (d).  The inset in panel (b) gives an enlarged view of the shaded region in a linear scale.\label{fig05}}
\end{figure*}

Figures \ref{fig05}(a) and (b) show the temporal evolutions of the looptop temperature and density derived from the HYDRAD simulations for $\Gamma=1$, $\Gamma=3$, and $\Gamma=10$, respectively. Similar to the patterns revealed in the analytical modeling, here the loop cross-sectional expansion also strongly suppresses the draining of the mass while not significantly affecting the evolution of the temperature. Besides the general consistency, some new features are also presented in the numerical simulations. First, the temperature cooling profiles exhibit a discernible dispersion. From $\Gamma=1$ to $\Gamma=10$, the loop cooling time changes from 2150 s to 2390 s, elongating by an amount of 240 s. Even so, the cooling time is still shorter than the analytically estimated value (2880 s). Second, for expanding loops, the apex density first rises, rather than immediately drops, during the early stage of the cooling. As illustrated in the inset of Figure \ref{fig05}(b), for $\Gamma=10$, the looptop density experiences an oscillatory increase, and keeps an elevated level until 1000 s after the start of the cooling.

The reason for these new features should lie in an inclusion of thermal conduction in the numerical modeling, which also acts as a cooling term in the corona. According to the initial equilibrium modeling, thermal conduction  plays a role comparable to or even more important than radiation during the early stage cooling \citep{Martens2010,Klimchuk2019}. A combination of conductive and radiative losses naturally leads to a shorter cooling time when comparing with the cooling by radiation alone. The effect of thermal conduction also depends on the loop geometry. Like the suppression of mass draining, an upward expanding (or a downward contracting) cross section blocks the downward transport of heat flux as well \citep{Antiochos1976}. Therefore, an expanding loop should cool somewhat more slowly than a uniform loop \citep{Antiochos1978,Johnson2024,Cargill2025}. On the other hand, the downward contraction of the cross section reduces the emitting volume in the lower part of the loop. When a heat flux is transported from above, even at a depressed level, the limited emitting material is still unable to convert it into radiation in time. As a result, the excess heat flux will drive an evaporative flow to fill the upper part of the loop, causing an early stage enhancement of the upper part density. 

As a complement, in Figures \ref{fig05}(c) and (d) we also plot the temporal evolutions of the average coronal temperature $T_c$ and density $n_c$ (for details on the calculation of the coronal averages, see Appendix \ref{appenda}), which show a close similarity to those of the looptop quantities. Such similarity can be understood in terms of a narrow variation of the temperature over the corona (within a factor of 2), and reflects a global effect of the cross-sectional expansion in influencing the loop thermodynamic evolution.

\begin{figure}
\epsscale{1}
\plotone{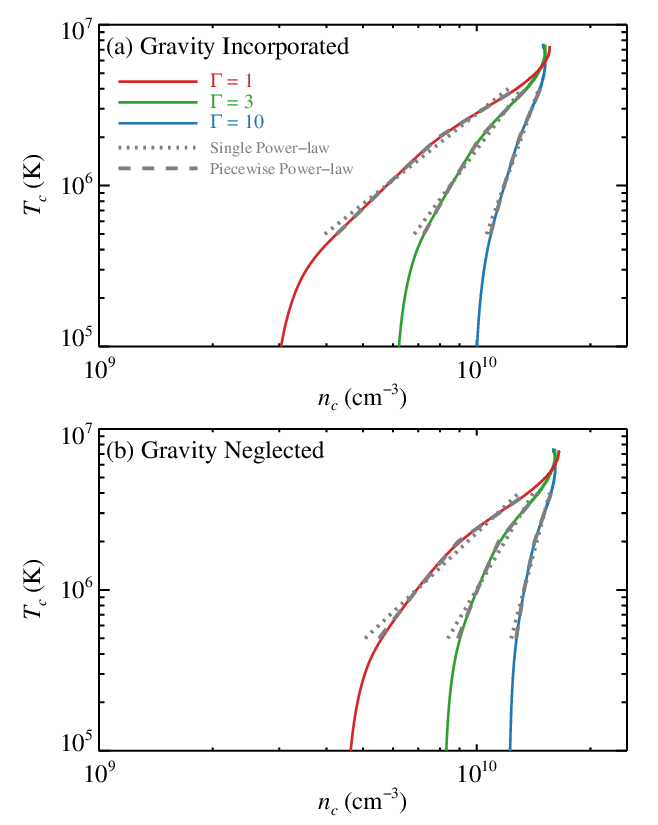}
\caption{Phase plots of the average coronal temperature vs. average coronal density based on the HYDRAD simulations, with different colors discriminating the cases of different values of $\Gamma$.  The upper panel shows the cases with gravitational stratification taken into account, while the bottom panel displays the cases with gravity neglected for comparison. In each panel, the dotted lines represent the results of a single power-law fit to the $T_c$--$n_c$ curves, whereas the dashed lines denote the piecewise power-law fit results. Both fits are performed within a temperature range of [4, 0.5] MK.\label{fig06}}
\end{figure}

\added{Based on the HYDRAD simulation results, we investigate how well the numerically modeled loop cooling process follows the analytically assumed $T$--$n$ scaling relation.  As mentioned above, the simulations presented in Figure \ref{fig05} are carried out with gravitational stratification taken into account for physical realism, which might cause an inconsistency when comparing with the analytical modeling, where the effect of gravity is neglected. To explore whether the incorporation of gravity might significantly affect the loop evolution, we have also performed another set of HYDRAD simulations in which the loop-aligned gravitational acceleration is switched off. The phase plots of the average coronal temperature versus average coronal density according to both sets of HYDRAD simulations are displayed in Figure \ref{fig06}.} Unlike the analytical solutions, the numerically derived $T_c$--$n_c$ curves exhibit an evolving slope that changes significantly during different stages of the cooling \citep{Bradshaw2010}. During the initial cooling stage (when $T_c>\sim5$ MK), the density barely decreases (for $\Gamma=1$) or even increases (for $\Gamma$=3 and 10). As mentioned above, thermal conduction still plays an important role in this stage. Its effect to drive an upward evaporation competes against the inclination of radiation to induce a downward draining, hence resulting in an extremely steep or even inverse slope of the $T_c$--$n_c$ curves. As thermal conduction gradually diminishes and radiation finally takes over, the loop then experiences a period of relatively steady cooling and draining, during which a valid $T$--$n$ scaling relation can be evaluated. At further lower temperatures (when $T_c<\sim$0.3 MK), the curves steepen again. According to \citet{Cargill2013}, in this regime, the sound speed has become too slow to maintain an effective communication between the TR and corona, without which the downward mass flow cannot adjust through sound waves in time to catch up with the fast temperature drop. Therefore, the scaling relation between $T$ and $n$ is broken again, and the loop cooling goes into a ``catastrophic" stage. 

\begin{deluxetable}{lcchcccc}
\tablecaption{Results of Power-law Fits to the Numerically Derived $T_c$--$n_c$ Curves\label{table02}}
\tablehead{
\colhead{} & \multicolumn{2}{c}{Single power-law fit} & \colhead{} & \multicolumn{4}{c}{Piecewise power-law fit}\\
\cline{2-3} \cline{5-8}
\colhead{} & \colhead{$\delta$ {\footnotesize ([4, 0.5] MK)}} & \colhead{$\chi^2$} & \colhead{} & \colhead{$\delta_1$ {\footnotesize ([4, 2] MK)}} & \colhead{$\delta_2$ {\footnotesize ([2, 1] MK)}} & \colhead{$\delta_3$ {\footnotesize ([1, 0.5] MK)}} & \colhead{$\chi^2$}
}
\startdata
\multicolumn{8}{c}{Gravity Incorporated}\\
$\Gamma=1$  & 1.86 & 475 & & 1.46 & 2.17 & 2.30 & 3.93 \\
$\Gamma=3$ & 3.18 & 186 & & 2.47 & 3.83 & 4.26 & 2.73 \\
$\Gamma=10$ & 6.52 & 19.3 & & 5.49 & 7.41 & 8.22 & 0.270\\
\hline
\multicolumn{8}{c}{Gravity Neglected}\\
$\Gamma=1$  & 2.22 & 555 & & 1.64 & 2.74 & 3.01 & 9.81 \\
$\Gamma=3$ & 4.00 & 202 & & 2.89 & 5.32 & 5.88 & 3.66 \\
$\Gamma=10$ & 8.88 & 25.2 & & 6.86 & 10.8 & 14.3 & 0.216
\enddata
\end{deluxetable}

We perform power-law fits to the $T_c$--$n_c$ curves to quantify the scaling relation. Here we adopt two approaches, namely, a single power-law fit to the curves within an overall temperature range of [4, 0.5] MK, and a piecewise power-law fit over three fixed subranges of  [4, 2], [2, 1], and [1, 0.5] MK evenly divided in logarithmic scale. The fit results are over-plotted in Figure \ref{fig06}, and the best-fit parameters are listed in Table \ref{table02}. For both approaches, it is evident that a larger expansion factor does lead to a higher value of the scaling index $\delta$, so the analytically inferred physical meaning of $\delta$ is now numerically consolidated. In spite of a general consistency between the two sets of fits, the piecewise power-law fit gives a better characterization of the loop cooling process, as reflected by a much closer match to each target $T_c$--$n_c$ curve (visually inspected in Figure \ref{fig06}), as well as a much smaller $\chi^2$ estimation of the fit (quantitatively compared in Table \ref{table02}). From the high temperature range to lower ones, the best-fit value of $\delta$ increases systematically (see Table \ref{table02}), implying a gradual steepening of the $T_c$--$n_c$ curves that eventually transitions to the final stage of catastrophic cooling. \added{Moreover, compared with the cases with gravity neglected, the incorporation of gravitational stratification leads to a systematical reduction of fitted $\delta$ (see Table \ref{table02}), which can be also visually inferred from the slope of the $T_c$--$n_c$ curves. Such a pattern can be understood in terms of a facilitation of downward mass motion (draining) by gravity \citep{Bradshaw2010}. In quantitative terms, nevertheless, the difference between the cases with and without gravity is not so prominent, except for quite low temperatures ($<$1 MK) at which the pressure scale height of the loop has dropped substantially. In this sense, the incorporation of gravitational stratification does not compromise the key signatures attributed to loop cross-sectional expansion, but just marginally influences their magnitudes. For this reason and for physical realism, we retain the cases with gravity in the following investigations.}


\begin{figure*}
\epsscale{1}
\plotone{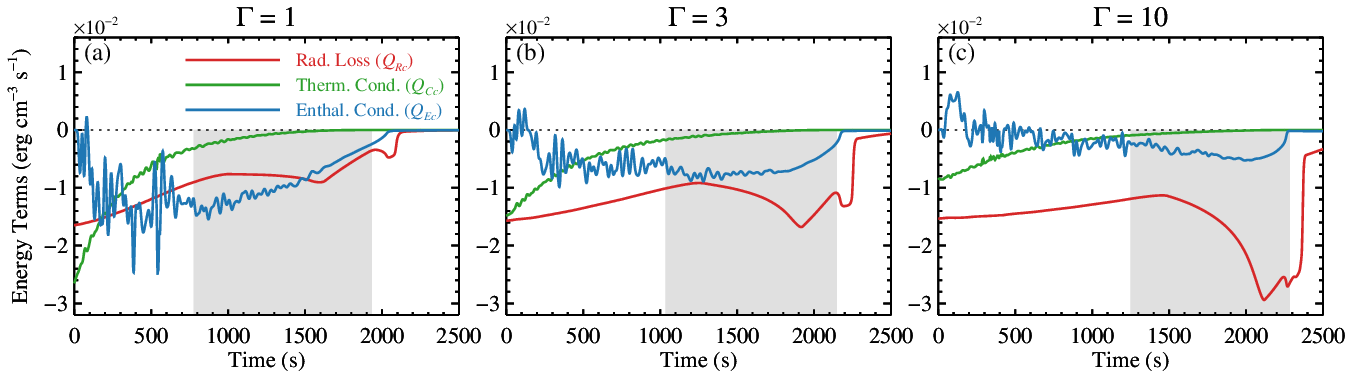}
\caption{Temporal evolutions of the average coronal radiation (red), thermal conduction (green), and enthalpy conduction (blue) based on the HYDRAD numerical simulations. Different panels display the cases of different $\Gamma$ values. In each panel, the shaded region highlights the time interval during which the average coronal temperature of the loop drops from 4~MK to 0.5~MK. \label{fig07}}
\end{figure*}

\added{The above power-law fits yield an overall value of fitted $\delta$ around 2 for case of $\Gamma=1$, $\sim$4 for $\Gamma=3$, and $\sim$8 for $\Gamma=10$, tentatively indicating a one-to-one physical correspondence between the three numerical cases presented here and the three analytical cases demonstrated in Section \ref{sec02}.} To verify such correspondence, we calculate the coronal averages of a variety of energy terms for the three numerical cases, and plot their temporal evolutions in Figure \ref{fig07}. With the incorporation of thermal conduction, enthalpy conduction is not always a cooling term. A persistent draining enthalpy flow is built up only when the thermal conduction is considerably weakened. Afterwards, the magnitudes of both enthalpy conduction and radiation decrease with time for $\Gamma=1$ (uniform cross section, Figure \ref{fig07}(a)), keep largely unchanged for $\Gamma=3$ (moderate cross-sectional expansion, Figure \ref{fig07}(b)), and increase instead for $\Gamma=10$ (strong cross-sectional expansion, Figure \ref{fig07}(c)), qualitatively consistent with the temporal trend for the three analytical cases revealed in Figure \ref{fig02}. Meanwhile, with the increase of $\Gamma$, it is seen that the magnitude of radiation increases while that of enthalpy conduction decreases, implying an increasing dominance of radiation.

\added{Quantitatively, we integrate the average coronal radiation and enthalpy conduction over a time interval when the average coronal temperature of the loop drops from 4~MK to 0.5~MK for each case (outlined by the shaded regions in Figure \ref{fig07}). The ratios of the time-integrated radiation to enthalpy conduction are 0.82, 1.71, and 4.54 for $\Gamma=$ 1, 3, and 10, respectively. These numerically calculated ratios are somewhat lower than the corresponding analytical values, which are 1.73, 3.00, and 5.53 for $\delta=$2, 4, and 8, respectively (see Figure \ref{fig03}). We attribute this discrepancy to several different model setups in the numerical simulations compared with the analytical modeling. First, we adopt a piecewise power-law formed radiative loss function in the numerical modeling, whose overall slope is steeper within the considered temperature range (indicative of a relatively lower coronal radiation). Second, the inclusion of gravitational stratification will further lead to a depletion of coronal density (and hence radiation) as well as an enhancement of mass draining (and hence enthalpy losses). Besides, the specification of the exact areal function form (even though the expansion factor is the same) and the definition of the coronal segment would also influence the ratio values. In spite of the notable quantitative difference, the analytically predicted effect of loop cross-sectional expansion to increase the dominance of radiation over enthalpy conduction, is still solidly retained in the numerical simulations.}

In passing, we also note from Figure \ref{fig07} that the term of enthalpy based on the numerical simulations exhibits notable fluctuations, in contrast to a smooth evolution in the analytical modeling. Such fluctuations may be due to a modulation of sound waves generated and bouncing to and fro within the loop, another piece of physics missing in the analytical consideration.  Moreover, as the expansion factor increases, the amplitude of the fluctuations decreases, implying a dampening of the sound waves by cross-sectional expansions \citep{Reep2022,Reep2024}. 

\begin{figure}
\epsscale{1}
\plotone{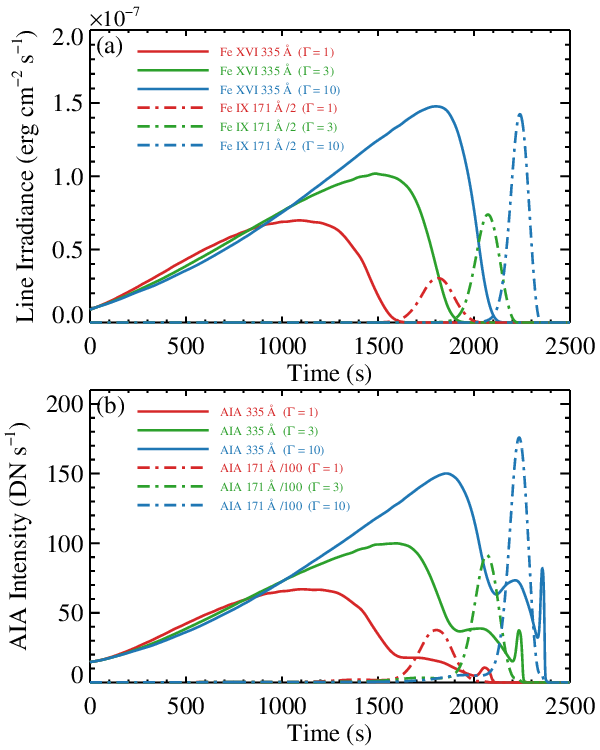}
\caption{The same as Figure \ref{fig04}, except that the light curves are synthesized with the numerical results for different values of $\Gamma$.\label{fig08}}
\end{figure}

\begin{deluxetable}{lccccchccccc}
\tablecaption{Quantities Relevant to the Peak of Synthetic Looptop Emissions for the Numerical Simulations\label{table03}}
\tablehead{
\colhead{} & \multicolumn{5}{c}{Peak of \ion{Fe}{16}/AIA 335 {\AA}} & \colhead{} & \multicolumn{5}{c}{Peak of \ion{Fe}{9}/AIA 171 {\AA}}\\
\cline{2-6} \cline{8-12}
\colhead{} & \colhead{$I_{\mathrm{peak}}$\tablenotemark{\scriptsize a}} & \colhead{$t_{\mathrm{peak}}$} & \colhead{$t_{\mathrm{decay}}$} & \colhead{$T_a$} & \colhead{$n_a$} & \colhead{} & \colhead{$I_{\mathrm{peak}}$\tablenotemark{\scriptsize a}} & \colhead{$t_{\mathrm{peak}}$}  & \colhead{$t_{\mathrm{decay}}$} & \colhead{$T_a$} & \colhead{$n_a$}\\
\colhead{} & \colhead{} & \colhead{\footnotesize (s)} &\colhead{\footnotesize (s)} & \colhead{\footnotesize(MK)} & \colhead{\footnotesize (cm$^{-3}$)} & \colhead{} & \colhead{} & \colhead{\footnotesize (s)} & \colhead{\footnotesize (s)} & \colhead{\footnotesize (MK)} & \colhead{\footnotesize(cm$^{-3}$)}
}
\startdata
& \multicolumn{11}{c}{Fe line irradiance}\\
$\Gamma=1$  & $6.98\times10^{-8}$  & 1091 & 353 & 3.29 & $9.67\times10^{9}$ &   & $6.06\times10^{-8}$ & 1813 & 93 & 0.849 & $4.55\times10^{9}$\\
$\Gamma=3$  & $1.02\times10^{-7}$  & 1486 & 292 & 3.01 & $1.11\times10^{10}$ &   & $1.48\times10^{-7}$ & 2073 & 69 & 0.837 & $7.08\times10^{9}$\\
$\Gamma=10$  & $1.48\times10^{-7}$  & 1802 & 202 & 2.77 & $1.29\times10^{10}$ &   & $2.85\times10^{-7}$ & 2239 & 52 & 0.831 & $9.82\times10^{9}$\\
\hline
& \multicolumn{11}{c}{AIA passband intensity}\\
$\Gamma=1$  & 67.0 & 1103 & 403 & 3.25 & $9.59\times10^{9}$ & & $3.76\times10^{3}$ & 1804 & 96 & 0.874 & $4.60\times10^{9}$\\
$\Gamma=3$  & 99.9  & 1594 & 252 & 2.70 & $1.05\times10^{10}$ & & $9.14\times10^{3}$ & 2067 & 71 & 0.862 & $7.14\times10^{9}$\\
$\Gamma=10$  & 150 & 1857 & 211 & 2.61 & $1.27\times10^{10}$ &  & $1.76\times10^{4}$ & 2235 & 53 & 0.852 & $9.87\times10^{9}$
\enddata
\tablenotetext{\scriptsize a}{\footnotesize Values are listed in units of erg cm$^{-2}$ s$^{-1}$ for the Fe line irradiance, or DN s$^{-1}$ for the AIA passband intensity.}
\end{deluxetable} 

Based on the numerical results, we finally synthesize loop emissions as done for the analytical solutions. The resultant light curves at the loop apex are plotted in Figure \ref{fig08}, and the quantities relevant to the synthetic emission peaks are summarized in Table \ref{table03}. As expected, a stronger loop cross-sectional expansion leads to a later occurrence of the emission peak in combination with a higher intensity, which is then followed by a sharper decay. In addition, because of the one-to-one correspondence, the peak-related quantities (except for the peak time) for each numerical case show a general consistency with the values for the corresponding analytical case.

\section{Observational Validation\label{sec04}}
\begin{figure*}
\epsscale{0.8}
\plotone{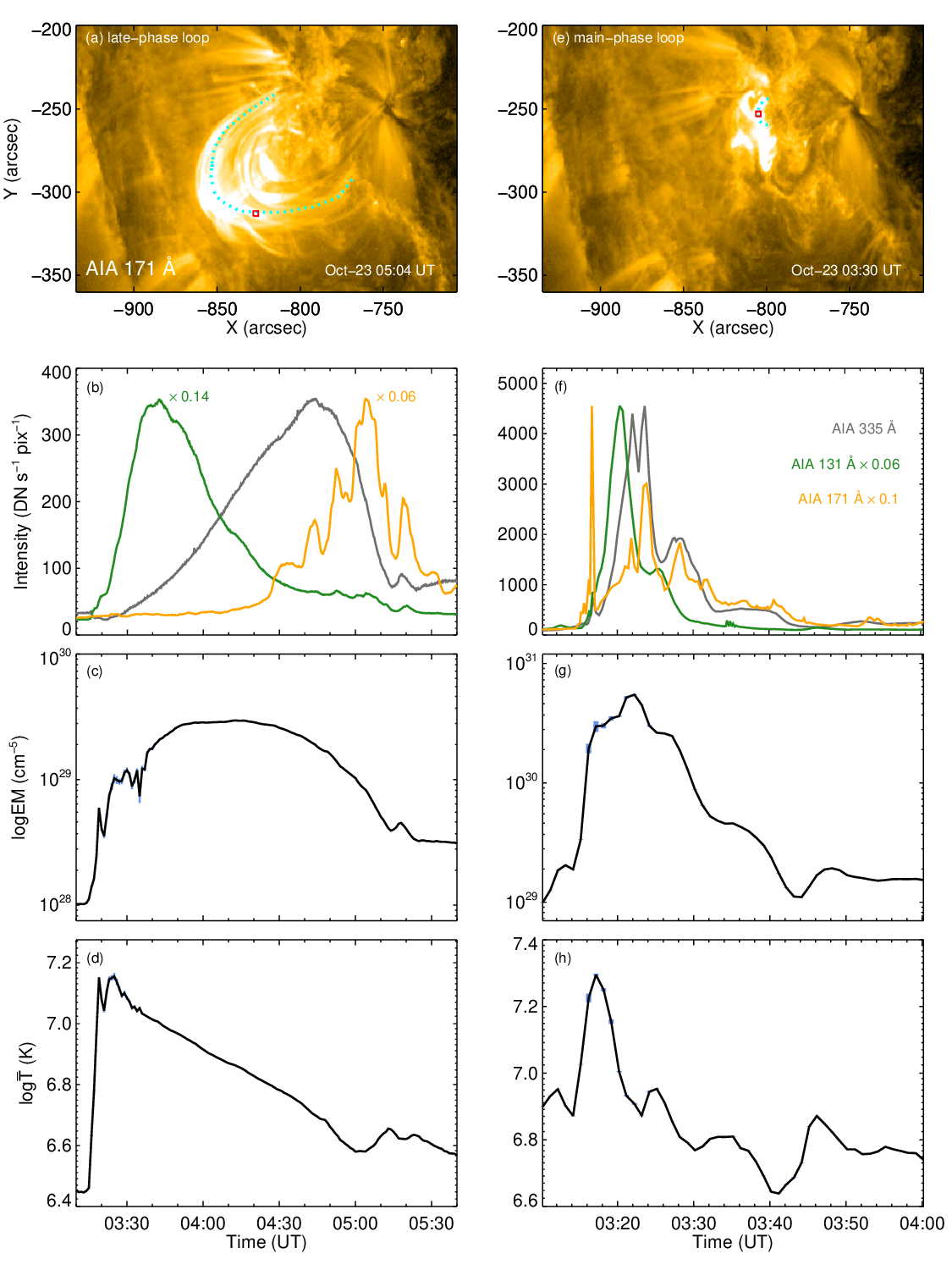}
\caption{Evolution of a representative late-phase loop (left) and main-phase loop (right) in the 2012 October 23 X1.8-class solar flare. In each column, the upper panel displays an AIA 171 {\AA} image of the flare region, where the dotted cyan curve outlines the spine of the representative flare loop, and the small red box (with a size of $3\arcsec\times3\arcsec$) encloses a close-to-apex region of the loop for further analysis, the middle-upper panel plots AIA light curves of the selected looptop region in three passbands (discriminated with different colors), and the lower two panels show temporal profiles of the column EM and DEM-weighted temperature based on DEM analysis to the same looptop region (with the light blue error bars indicating the errors of both quantities).\label{fig09}}
\end{figure*}

To validate the analytically and numerically predicted effect of cross-sectional expansion from an observational perspective, we turn to EUV observations of an extensively studied X1.8-class solar flare occurring on 2012 October 23 \citep{Yang2015,Watanabe2020,Wu2023,Liu2024,Li2024}. The flare exhibits an extremely large late-phase peak in the warm coronal emissions (e.g., in AIA 335~{\AA}), which originates from a set of longer and higher flare loops other than the main-phase flare loops (see Figure 2 in \citealt{Li2024}). Figure \ref{fig09}(a) displays the late-phase loops when they become visible in the cooler AIA 171~{\AA} passband, from which we trace out a representative loop (outlined by the dotted cyan curve). \added{Using a Gaussian fit to the cross-spine AIA intensity profiles, \citet{Li2024} tracked the variation of cross section along this loop, finding an expansion of 33\% over a loop segment where the loop is clearly distinguishable from other loops (Figure 6 in their paper). In addition, magnetic modeling based on force-free field extrapolations also revealed a decay of the loop-hosting magnetic field that is quantitatively compatible with the cross-sectional expansion along the selected segment.}

\added{For this representative late-phase loop, we select a $3\arcsec\times3\arcsec$ sized region near the loop apex (enclosed by the small red box), and calculate the mean intensities of this region in three AIA passbands (including the hot coronal AIA 131~{\AA} passband in addition to AIA 335 and 171 {\AA}), whose light curves are plotted in Figure \ref{fig09}(b).} First, a sequential occurrence of the emission peaks in decreasing temperatures suggests a general cooling process of the late-phase loop. More importantly, the AIA 335~{\AA} intensity profile shows a fairly sharp decay preceded by a rather gradual rise, consistent with the emission pattern of an expanding loop revealed in the synthetic light curves (Figures \ref{fig04} and \ref{fig08}).

For comparison, we also pick out a representative main-phase loop (Figure \ref{fig09}(e)), and plot the AIA light curves of the looptop region in Figure \ref{fig09}(f). Due to its much shorter length,  the main-phase loop cools much faster than the late-phase loop. Meanwhile, it is unlikely that a notable cross-sectional expansion could develop within such a short loop length. Reflected to the shape of the AIA 335 {\AA} light curve,  the duration of its decay phase does not show a remarkable difference with the rise phase.

Expanding the AIA observations to all six optically thin passbands, we conduct differential emission measure (DEM) analysis on the selected loop regions. The DEM function is defined as
\begin{equation}
\mathrm{DEM}(T)=n^2(z)\frac{dz}{dT},
\end{equation}
with $z$ being the distance along the LOS, and its distribution can be evaluated by solving the integral equations 
\begin{equation}
I_{\mathrm{AIA}i}=\int K_i(T)\times\mathrm{DEM}(T)dT,
\end{equation}
where the subscript $i$ denotes the $i$th AIA passband included for analysis. Using the sparse inversion algorithm \citep{Cheung2015,Su2018}, we construct the DEM distributions over a total of 42 temperature bins that are evenly divided in logarithmic scale within an overall range of $\log(T/\mathrm{K})\in[5.5,\ 7.6]$. \added{In practice, we perform a total of 100 Monte Carlo runs for each DEM inversion, adopting the mean of the inversion results in each temperature bin as the resultant DEM and the standard deviation of them as the inversion error.}

Based on the DEM inversion results, we compute the column emission measure (EM)
\begin{equation}
\mathrm{EM}=\int n^2(z) dz=\int \mathrm{DEM}(T) dT,\label{cem}
\end{equation}
and the DEM-weighted temperature
\begin{equation}
\bar{T}=\frac{\int T\times\mathrm{DEM}(T) dT}{\mathrm{EM}}.
\end{equation}
\added{The errors of these two quantities are then estimated using the error propagation formula.} Figures \ref{fig09}(c) and (d) display the temporal evolution of EM and $\bar{T}$ for the selected late-phase loop region, and Figures \ref{fig09}(g) and (h) show that for the main-phase loop region. \added{Note that the errors for both parameters are very small (barely discernible at most moments as shown in the figure), which corroborates the robustness of the DEM inversion algorithm.} As both loops cool down, the EM (closely related to density according to Equation (\ref{cem})) of the late-phase loop maintains a plateau level for a long time with just a slight drop, while such a plateau is absent in the EM curve of the main-phase loop. \added{Besides, in Appendix \ref{appendb} we complement the same analysis for two additional late-phase loop regions, which reveals a similar trend for the AIA light curves and EM evolution (Figure \ref{fig10}).} In this sense, the suppression of mass draining by cross-sectional expansion is observationally consolidated for this late-phase flare. 

\section{Discussion and Conclusions\label{sec05}}
In this work, we carry out both analytical and numerical loop-aligned hydrodynamic modelings of the radiative cooling in a flare loop. The derivation of the analytical solutions is based on the assumption of a temperature--density scaling relation of $T\sim n^{\delta}$. It is seen that a higher value of the scaling index $\delta$ corresponds to a less prominent mass draining from the corona (see Figure \ref{fig01}), which can be physically realized by an upward expansion of the loop cross section. In the following numerical simulations that explicitly parameterize the variations of loop cross section, this argument is further consolidated, where an increase of the expansion factor $\Gamma$ obviously steepens the $T$--$n$ phase curve (see Figure \ref{fig06}), consistent with the findings in other numerical studies \citep{Reep2022,Reep2024}.

According to our modelings, the cross-sectional expansion has a profound influence on the loop energetics. First, by directly reducing the enthalpy losses from the corona (owing to a lowered energy requirement to power the TR radiation from a contracted volume), the coronal part of the loop will cool more dominantly by radiation (see Figures \ref{fig03} and \ref{fig07}). More importantly, the partition of the radiative outputs between different temperatures is essentially changed. During the radiative cooling stage, the evolution of radiation is the result of a competition between the decrease of density and the increase of radiative loss function ($\Lambda(T)\sim T^{-l}$). If the loop material is drained significantly (as in the case of a uniform loop), the effect of density always dominates so that the radiation decreases with time (as well as temperature). However, by efficiently suppressing the mass draining (as in the case of an expanding loop), the effect of radiation function wins out instead, hence leading to a transition of the evolution trend, namely, a shift of the radiative outputs toward lower temperatures (see Figures \ref{fig02} and \ref{fig07}). 

Such a transition may pose important physical implications for coronal emissions observed from a flare loop. Regarding to EUV late-phase flares \citep{Woods2011}, the late-phase loops in these flares are believed to bear a more notable cross-sectional expansion owing to their longer lengths \citep{Liu2013,Dai2013,Sun2013,Masson2017,Li2024}, a natural consequence of the decay of coronal magnetic fields with height \citep{Asgari-Targhi2012}. Compared with the main-phase loops, therefore, the late-phase loops would emit more effectively at middle temperatures (see Figures \ref{fig04} and \ref{fig08}), which could, to a certain degree, mitigate the severe heating requirement for the production of a prominent warm coronal  late-phase peak. Of course, the cross-sectional expansion can give rise to an even more pronounced enhancement of the cool coronal emissions as observed in the \ion{Fe}{9} 171 {\AA} line or AIA 171 {\AA} passband. Nevertheless, the emissions at these temperatures are contributed dominantly by the bulk corona. Background fluctuations during the flare interval and/or coronal dimmings induced by an associated CME \citep{Aschwanden2009,Cheng2016} can easily submerge the cool coronal late phase.

In addition to the emission intensity, the cross-sectional expansion also affects the shape of the emission light curves. By sustaining the material initially evaporated to the corona for a long time, the warm (and cool) coronal emission peak in an expanding loop occurs closer to the maximum of the corresponding contribution or temperature response function (see Tables \ref{table01} and \ref{table03}). Afterwards, the decreases of both loop density and contribution/response function will result in a sharper decay of the emission. Such an emission pattern has been validated with the observations of an EUV late-phase flare (see Figures \ref{fig09} \added{and \ref{fig10}}), and \added{could} serve as a \added{potential} diagnostic tool to judge the degree of loop cross-sectional expansion in an extended flare dataset.

Due to its profound influence on loop evolution, pointed out here and suggested in other studies \citep{Mikic2013,Reep2023}, we emphasize the importance of including cross-sectional expansion in sophisticated loop-aligned hydrodynamic modelings. An important issue is that how to quantitatively determine the cross-sectional expansion from an observational basis. Although magnetic modelings usually imply a large cross-sectional expansion in coronal loops \citep{Asgari-Targhi2012,Chen2022}, observations typically show a nearly constant cross section along most coronal loops \citep{Klimchuk2000,Klimchuk2020}. Exploring the underlying reason for this apparent contradiction is out the scope of this work. We hope that the improved spatial resolution and temperature coverage provided by current and future solar instruments can help clarify this issue \citep{Williams2021,Tamburri2025}.

\begin{acknowledgments}
\added{We are grateful to the anonymous referee for his/her insightful comments and suggestions.} This work was supported by the National Natural Science Foundation of China under grant 12127901, and the Fundamental Research Funds for the Central Universities under grant KG202506. \emph{SDO} is a mission of NASA's Living With a Star (LWS) program.
\end{acknowledgments}

\appendix
\section{Calculation of Average Quantities over the Corona}\label{appenda}
Considering a variation of the loop cross section, the average of a loop quantity over the corona can be defined as
\begin{equation}
\psi_c=\frac{\int_{s_1}^{s_2}A\psi ds}{\int_{s_1}^{s_2}A ds},
\end{equation}
where $s_1$ and $s_2$ are the coordinates of two opposite coronal footpoints of the loop, $\psi$ is the loop quantity to be averaged, and $\psi_c$ is its area-weighted average between $s_1$ and $s_2$. In this work, we assume a corona extending for 80\% of the loop full-length (with two chromospheric feet excluded), by which $s_1=12$~Mm and $s_2=92$~Mm.

For the field-aligned conductive terms of thermal conduction $Q_C=-(1/A)\partial(AF_C)/\partial s$ and enthalpy conduction $Q_E=-(1/A)\partial(AF_E)/\partial s$, the average calculation is mathematically more straightforward, which yields
\begin{equation}
Q_{Cc}=-\frac{\left.(AF_C)\right|_{s=s_1}^{s=s_2}}{\int_{s_1}^{s_2}A ds}
\end{equation} 
and
\begin{equation}
Q_{Ec}=-\frac{\left.(AF_E)\right|_{s=s_1}^{s=s_2}}{\int_{s_1}^{s_2}A ds}.
\end{equation} 

\added{\section{Analysis for Additional Late-phase Loops}\label{appendb}
The observational validation of the suppression of mass draining by cross-sectional expansion presented in Section \ref{sec04} is based on the analysis of a representative late-phase loop region. To test the robustness of our argument, we pick up two additional late-phase loop regions that are well separated from the former representative one, and perform the same analysis. The results for these two regions are displayed in Figure \ref{fig10}, which reveals a similar trend for the AIA light curves and EM evolution. It is worth pointing out that \citet{Li2024} calculated the total EM over the whole late-phase loop region for this flare, whose evolution is also consistent with that of the subregions in this work (Figure 4 in their paper). Therefore, the signature revealed from the DEM analysis should represent a global feature for a large sample of late-phase loops, largely insensitive to manual region selection.
\begin{figure*}
\epsscale{0.8}
\plotone{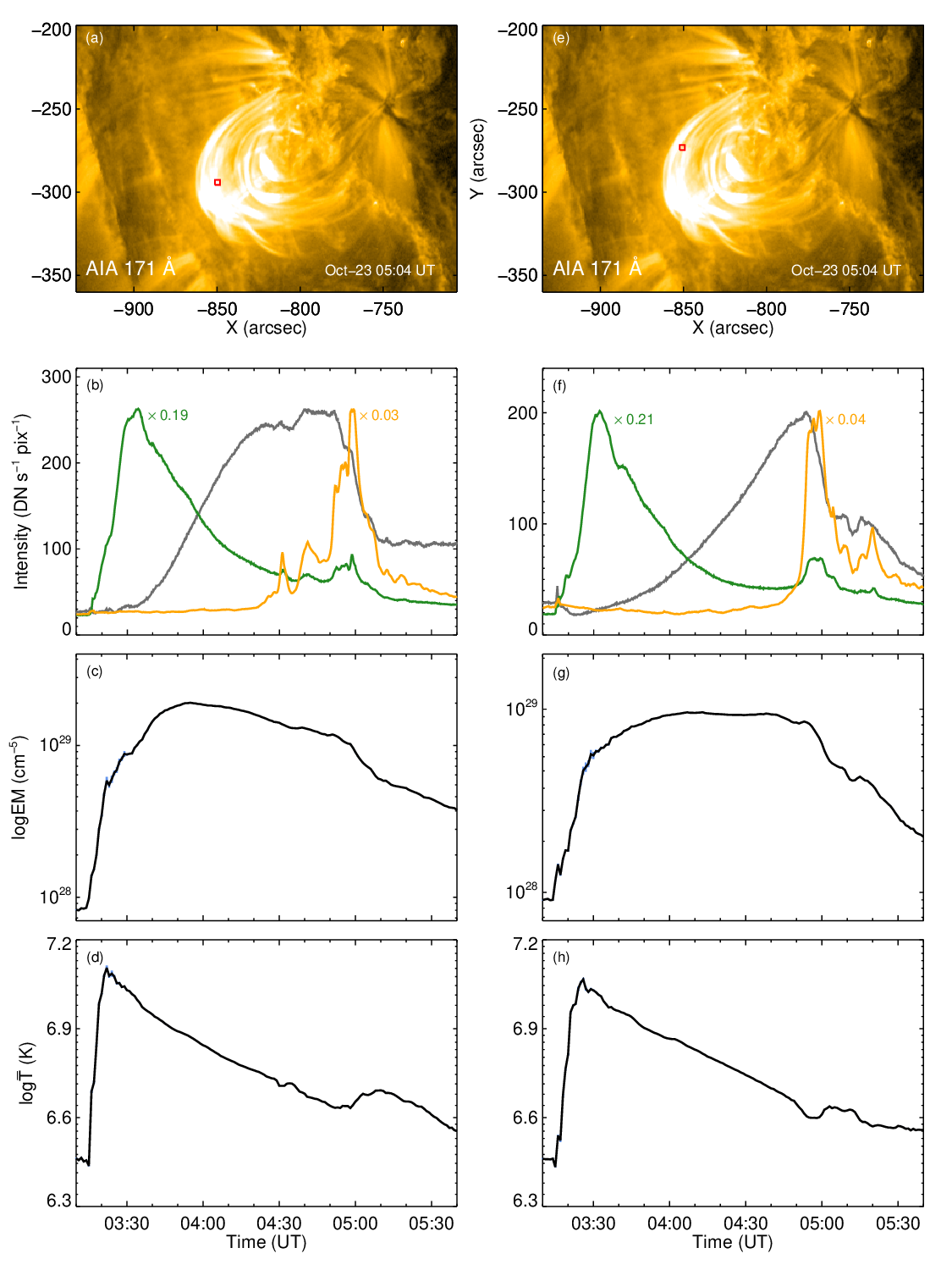}
\caption{Evolution of two additionally selected late-phase loops in the 2012 October 23 X1.8-class solar flare. The panels are organized the same as in Figure \ref{fig09}.\label{fig10}}
\end{figure*}
}


\end{document}